# Ultracompact Vanadium Dioxide Dual-Mode Plasmonic Waveguide Electroabsorption Modulator

**K J A Ooi[1,2], P Bai[1,*], H S Chu[1] and L K Ang[2,3]**

[1]Electronics & Photonics Department, A*STAR Institute of High Performance Computing, 1 Fusionopolis Way, #16-16 Connexis, Singapore 138632

[2]School of Electrical and Electronic Engineering, Nanyang Technological University, Block S1, 50 Nanyang Avenue, Singapore 639798

[3]Engineering Product Development, Singapore University of Technology and Design, 20 Dover Drive, Singapore 138682

*Email: baiping@ihpc.a-star.edu.sg

**Abstract**

Subwavelength modulators play an indispensable role in integrated photonic-electronic circuits. Due to weak light-matter interactions, it is always a challenge to develop a modulator with a nanometer scale footprint, low switching energy, low insertion loss and large modulation depth. In this paper, we propose the design of a vanadium dioxide dual-mode plasmonic waveguide electroabsorption modulator using a metal-insulator-$VO_2$-insulator-metal (MIVIM) waveguide platform. By varying the index of vanadium dioxide, the modulator can route plasmonic waves through the low-loss dielectric insulator layer during the "on" state and high-loss $VO_2$ layer during the "off" state, thereby significantly reducing the insertion loss while maintaining a large modulation depth. This ultracompact waveguide modulator, for example, can achieve a large modulation depth of ~10dB with an active size of only $200\times50\times220$nm$^3$ (or ~$\lambda^3/1700$), requiring a drive-voltage of ~4.6V. This high performance plasmonic modulator could potentially be one of the keys towards fully-integrated plasmonic nanocircuits in the next-generation chip technology.

**1 Introduction**

The continuous scaling down in electronic chips has resulted in the electrical interconnect bottleneck problem [1]. One solution is to replace the electrical interconnect with a silicon photonic waveguide, which is superior in terms of negligible propagation loss, high transmission speed and huge data-carrying capacity [2]. However, photonic-based optoelectronic devices suffer from low response times and high energy consumption due to their sizes. Hence, we look towards plasmonics as a possible device technology to overcome the problems faced by photonic devices. The nature of surface plasmons that allows light to be squeezed far below the diffraction limit has generated interest in the development of miniature chip-scale optoelectronic devices that promise high responsivity and low energy consumption [3]. Progress is seen in the development of plasmonic waveguides [4–6], plasmonic photodetectors [7–9], and plasmonic modulators [10–14].

Plasmonic modulators have numerous performance requirements to meet, such as small device footprint, low switching energy consumption, low insertion loss, high switching speed and large modulation depth [10]. These performance indicators are heavily influenced by both device structure design and the nonlinear coefficient of the active material. Robust structure design could reduce device insertion losses and switching energies. However, the modulation depth – the crucial performance parameter for modulators – is largely determined by the nonlinear coefficient of the active material. For example, silicon, the favored CMOS compatible material, only records a refractive index change, $\Delta n$ of $\sim 10^{-3}$ [15]. Barium titanate, meanwhile, can show $\Delta n$ as high as $\sim 0.05$ [16]. Recently, indium tin oxide (ITO) garnered some research interest due to its ability to show unity-order index change [17, 18].

Here, we discuss a special class of materials which can show similar unity-order index change but occurs throughout the bulk material – the Mott insulator. The Mott insulator is a solid state phase change material which shows a first order insulator-metal transition (IMT) under certain conditions [19], and it is usually accompanied by a large refractive index change. Vanadium dioxide ($VO_2$) is one of the Mott insulator candidates, which shows a large refractive index contrast of n=3.243+0.3466i for the insulating phase to n=1.977+2.53i for the metallic phase at the operating wavelength of 1550nm [20]. The phase transition can be triggered by thermal excitation [21], optical excitation [22], electrical excitation [23], doping [24], or strain engineering [25]. For an electrical excitation scheme, the field required for the phase change to occur is experimentally recorded to be $6.5 \times 10^7$ V/m [23]. Independent experiments show that the phase transition is typically picoseconds or faster, making it a potentially ultrafast switching material [26]. Hence, $VO_2$ has recently attracted research interest in various optical modulator designs [27, 28]. In [27], the authors achieved 6.5dB modulation with 2dB insertion loss for a 2µm long device and 65nm thin $VO_2$ films, using a thermal-excitation scheme. In [28], 20dB modulation is achieved, however with a thick 1µm waveguide stack.

In this paper, we propose a new type of $VO_2$ modulator by using a "dual-mode" plasmonic waveguide to realize a low-loss, low energy consumption and highly-compact plasmonic modulator. Taking advantage of the large refractive index contrast between the insulating and metallic phases of $VO_2$, the dual-mode plasmonic waveguide modulator has the ability to switch transmission modes between a low loss hybrid plasmonic mode and a high loss metal-insulator-metal (MIM) mode during the "on" and "off" states respectively. Thus, the modulator is able to achieve very high modulation depth (~10dB) with low insertion loss (~1dB) with this novel modulation scheme.

**2 Vanadium dioxide plasmonic slot waveguide modulator**

To evaluate the switching performance of $VO_2$, we first consider a simple plasmonic slot modulator using $VO_2$ as its active material. The $VO_2$ is sandwiched between two copper layers to form a metal-insulator-metal (MIM) waveguide as shown in Figure 1(A). We used copper (Cu, ε=−122+6.2i) as the plasmonic material in building the waveguide due to its low propagation loss among CMOS compatible metal materials in the near-infrared wavelengths [29]. The front and back ends of the $VO_2$ slot is interfaced with copper-silicon-copper MIM plasmonic tapered couplers. Silicon taper (n=3.48) is used to index-match with the insulating phase of $VO_2$ to minimize coupling loss. The plasmonic tapered couplers have widths decreasing linearly from 400nm to the slot width that couple light to and from 400nm wide silicon waveguides. The whole device is planar and 220nm thick. Since this device will be used as an electroabsorption modulator, we denote the $VO_2$ insulating phase as the "on" state, and the metallic phase as the "off" state due to its increased extinction coefficient.

We simulated this device using C.S.T. Microwave Studio 2012 [30]. Figure 1(B) shows the electric-field of the surface plasmons being confined strongly inside the Cu–VO$_2$–Cu slot, observed at the cross-sectional cutting plane. Next, Figure 1(C) shows the variation of the insertion loss and modulation depth with slot widths. At ultra-narrow slot widths (10nm–40nm), a tight confinement in the MIM waveguide results in high propagation losses. The loss does not reduce further for slot widths ≥50nm, and hence we choose 50nm for the slot width to obtain a relative low insertion loss and a reasonable modulation-voltage. Meanwhile, Figure 1(D) shows the insertion loss (optical attenuation in dB units in the "on" state) and modulation depth (optical attenuation in dB units in the "off" state less the insertion loss) for a 50nm wide slot waveguide modulator, measured to be around 10dB/μm and 50dB/μm respectively. For a 200nm long waveguide modulator, it would have a 2dB insertion loss and a 10dB modulation depth. Optimized couplers have a ~0.5dB loss per facet. At an electric-field threshold of ~6.5×10$^7$ V/m [23], the device requires a drive-voltage of ~3.3V arising from the narrow slot width of 50nm.

Energy consumption of the device is roughly divided into three components: capacitive loss across the VO$_2$ waveguide, capacitive loss across the plasmonic tapers, and joule heating loss due to leakage currents. If the modulator is assumed to be a simple capacitive device, the 200nm waveguide modulator would require an energy/bit of $\frac{1}{2}CV^2 = \frac{1}{2}\frac{\varepsilon\varepsilon_0 A}{d}V^2 = 1.53\text{fJ/bit}$ (using $\varepsilon_{VO2\text{-}I}$ (constant) ~ 36 [31], $\varepsilon_0 = 8.85\text{e}^{-12}$ Fm$^{-1}$). For the silicon plasmonic tapers, if we assume that they are 200nm in length and have an average width of 225nm, the energy/bit loss of each taper is 0.113 fJ/bit (using $\varepsilon_{Si}$ (constant) ~ 12). Finally, the joule heating term, given as $\frac{V^2}{RB} = \frac{V^2}{\left(\frac{\rho d}{A}\right)B} = 9.6\text{fJ/bit}$ (using $\rho_{VO2\text{-}M}$ ~ 1Ωm [32], assuming B = 1 GHz operating bandwidth). The total energy consumption is 11.4 fJ/bit, with the substantial waste coming from the joule heating loss.

Despite the exceptional switching performance of VO$_2$, it suffers from high insertion loss due to the plasmonic mode being concentrated in the lossy VO$_2$ medium. To reduce the loss, one idea is to reroute the surface plasmons from the VO$_2$ layer to another lossless or low-loss dielectric layer during the "on" state, since we do not require the surface plasmons to be present in the insulating phase of VO$_2$. This idea leads us to the design of a dual-mode plasmonic waveguide modulator, which will be discussed in the next section.

**3 Vanadium dioxide dual-mode plasmonic waveguide modulator**

*3.1 Dual-mode waveguide design*

Here, we design a VO$_2$ waveguide modulator with two transmission modes that can be switched by varying the index of the VO$_2$. Figure 2(A) shows the schematic of the proposed plasmonic waveguide modulator. The device is similar to the plasmonic slot waveguide modulator shown in Figure 1(A), except that the VO$_2$ and silicon plasmonic tapered couplers are now coated with a thin layer of lossless dielectric before being flanked by the metal layers, forming a metal-insulator-VO$_2$-insulator-metal (MIVIM) waveguide. During the "off" state, the surface plasmons undergo high extinction in the metallic VO$_2$ layer, and thus the electric-field distribution is similar to the MIM mode shown in Figure 1(B). During the "on" state, we switch the transmission to the low-loss cladding layers to reduce the insertion loss. This is done by raising the refractive index of VO$_2$ and converting the electric-field distribution to a hybrid plasmonic (MIVIM) mode as shown in Figure 2(B) [4, 5]. We refer to this design as a "dual-mode waveguide", the primary mode transmitting through the VO$_2$ core layer, and the secondary mode through the cladding layers of lossless dielectric. The structure of this waveguide is in-part inspired by recent-fabricated multi-layer nanoplasmonic slot waveguides which are easy to fabricate [33, 34].

The refractive index of the dielectric slot is important in the mode switching mechanism of the dual-mode waveguide, and hence affecting the modulation performance. Due to the large refractive index difference between the insulator (VO$_2$-I) and metallic (VO$_2$-M) phases of VO$_2$, there are three distinct categories of dielectric slot refractive indices: $n_{slot} < n_{VO_2-M}$, $n_{VO_2-M} < n_{slot} < n_{VO_2-I}$ and $n_{VO_2-I} < n_{slot}$. The proportion of surface plasmon modes travelling in the dielectric and VO$_2$ layers would depend on the category in which the dielectric slot refractive index belongs to, and hence determine the surface plasmon mode profile. In general, a higher portion of surface plasmons would travel in a lower index region. Hence, in both the "on" and "off" states, the $n_{slot} < n_{VO_2-M}$ modulator operates in the hybrid plasmonic mode, while the $n_{VO_2-I} < n_{slot}$ modulator operates in the MIM mode. However, in the case of $n_{VO_2-M} < n_{slot} < n_{VO_2-I}$, the modulator operates in the hybrid plasmonic mode only in the "on" state. In the "off" state, when the metallic VO$_2$ index is lower than that of the dielectric slot, the modulator switches to the MIM mode. We specially refer to this property as "modal switching".

The benefit of the modal switching is clearly seen in Figure 3. On this figure's left axis, we see two clearly distinct trends, in which the insertion loss increases monolithically with the dielectric refractive index, while the modulation depth increases in the form of a sigmoid curve. This modal switching causes a large appreciation of modulation depth with refractive index in the $n_{VO_2-M} < n_{slot} < n_{VO_2-I}$ region. This is very useful to maintain low insertion losses while gaining reasonably high modulation depths. On the right axis of the figure, the modulation depth to insertion loss ratio is plotted. This ratio is considered as a useful figure-of-merit to determine the optimal dielectric slot index for the modulator to operate in, whereby the drop in the insertion loss should over-compensate the drop in the modulation depth. It is observed that this ratio peaks at refractive indices in the $n_{VO_2-M} < n_{slot} < n_{VO_2-I}$ region, hence confirming that the best modulation performance occurs through the modal switching.

*3.2 Layer geometry*

While it was mentioned above that the refractive indices of the dielectric slot and VO$_2$ determine the proportion of travelling surface plasmon modes, here the slot widths would determine the "carrying capacity" of the slot layers. That is, on top of distribution according to the refractive index, the surface plasmon modes also distribute according to the dimensions. For example, in a Cu-SiO$_2$-VO$_2$-SiO$_2$-Cu waveguide stack, variation of the widths of the SiO$_2$ and VO$_2$ layers will affect the insertion loss as well as the modulation depth. Figure 2(C) shows that in general, the insertion loss increases with increasing VO$_2$ slot widths and decreasing SiO$_2$ slot widths. Similar trends could be seen for the modulation depth as well in Figure 2(D). The choice of dielectric and VO$_2$ slot widths should therefore consider the tradeoff between low insertion losses, high modulation depths and low drive-voltages to obtain an optimal device performance. For the subsequent discussions we will use a VO$_2$ slot width of 50nm and a dielectric slot width of 10nm as our reference dimensions.

Table 1. Three dielectric slot materials and their relationships with the two phases of VO$_2$

|  | VO$_2$ insulator phase (VO$_2$-I), n=3.243+0.3466i  |  |
|---|---|---|
|  | VO$_2$ metallic phase (VO$_2$-M), n=1.977+2.53i |  |
| SiO$_2$, n=1.44 | TiO$_2$, n=2.7 | Ge, n=4.27 |
| $n_{SiO_2} < n_{VO_2-M}$ | $n_{VO_2-M} < n_{TiO_2} < n_{VO_2-I}$ | $n_{VO_2-I} < n_{Ge}$ |

*3.3 Case studies*

Here we examine three different dielectric slot materials (SiO$_2$, TiO$_2$ and Ge), characterized by their relationships with the two phases of VO$_2$ as tabulated in Table 1. Figure 4 shows the simulation results for the three different dielectric slot materials. For the case of SiO$_2$ in Figure 4(A), the insertion loss is 2dB/μm, and modulation depth is 13dB/μm. For TiO$_2$ in Figure 4(B), the insertion loss is 5dB/μm, and the modulation depth is 45dB/μm. For Ge in Figure 4(C), the modulation depth is the highest at 57dB/μm, but comes with a larger insertion loss of 8dB/μm. Optimized coupling losses from the plasmonic tapered couplers are determined to be ~0.5dB per facet on average.

Figures 4(D), (E) and (F) are the index-normalized lateral electric-field profiles of the dual-mode waveguide for SiO$_2$, TiO$_2$ and Ge respectively. For SiO$_2$, in both "on" and "off" cases, the modulator operates in the hybrid plasmonic mode, hence explaining the low insertion loss and low modulation depth. For Ge, in both cases the modulator operates in the MIM mode, hence the high insertion loss and modulation depth. However, for TiO$_2$, the modal switching occurring between the "on" and "off" states are remarkably seen in Figure 4(E). In the "on" state, the modulator operates in the hybrid plasmonic mode, confining the surface plasmons in the dielectric slot. In the "off" state, the modulator operates in the MIM mode and as a result a great portion of the surface plasmons is transmitted in the VO$_2$ layer.

Table 2. Modulation depth to insertion loss ratio of various VO$_2$ dual-mode plasmonic waveguide modulators (200nm length)

|  | **Dielectric slot material** | | | |
|---|---|---|---|---|
|  | No slot | SiO$_2$ | TiO$_2$ | Ge |
| Insertion Loss | 2dB | 0.4dB | 1dB | 1.6dB |
| Modulation Depth | 10dB | 2.6dB | 9dB | 11.4dB |
| Ratio | 5 | 6.5 | 9.0 | 7.1 |

Table 2 shows the modulation depth to insertion loss ratio of various insulating slot materials of 200nm length VO$_2$ dual-mode plasmonic waveguide modulators. The "no slot" column refers to the simple plasmonic slot waveguide modulator in section 2. From the ratio figures, it is evident that TiO$_2$ shows the largest ratio of modulation depth and insertion loss. This analysis validates TiO$_2$ to be the optimal dielectric material choice among the three reported materials.

The additional 10nm dielectric slot layers slightly increases the waveguide width to 70nm. Using an electrical excitation scheme with a field threshold of ~$6.5\times10^7$ V/m, we work out the drive-voltage to be ~4.6V. The capacitive loss of the waveguide is 2.12fJ/bit, and 0.220fJ/bit for each silicon plasmonic taper. Leakage currents which contribute to joule heating loss will be negligible. Therefore, the total energy consumption is very much reduced to 2.6fJ/bit.

Finally, we compare our results with a recent work using similar modal-switching concept for a hybrid $VO_2$ plasmonic switch, but employed a metal-insulator-$VO_2$ (MIV) structure [35]. While the MIV structure efficiently reduced the insertion loss of the device during the "on" state, it suffered from low modulation depth because the surface plasmons are not efficiently confined in the $VO_2$ layer during the "off" state. The MIV design in [35] reported a highest modulation depth of only 6.1dB/μm.

**4 Conclusion**

We have proposed an ultracompact, high modulation depth, low insertion loss, and low energy consumption plasmonic modulator by employing $VO_2$ in a plasmonic dual-mode waveguide. The dual-mode configuration has the ability to route surface plasmons through low-loss or lossless dielectric layers in the "on" state, and rerouting them through the lossy $VO_2$ medium in the "off" state, thereby significantly reducing insertion loss while maintaining a high modulation depth. The designed modulator could be as short as ~200nm (~λ/8) in length, and have low insertion loss (~1dB) and high modulation depth (~10dB). The device's drive-voltage is ~4.6V, with energy/bit as low as 2.6fJ/bit. This high performance modulator contributes a step closer to realizing fully-integrated nanophotonic-nanoelectronic nanocircuits in next-generation chip technology.

**Acknowledgments**

This work was supported by the Agency for Science and Technology Research (A*STAR), Singapore, Metamaterials-Nanoplasmonics research programme under A*STAR-SERC grant No. 0921540098. KJAO is supported by a PhD scholarship funded by the MOE Tier2 grant (2008-T2-01-033).

**References**

1. M. T. Bohr, "Interconnect scaling – The real limiter to high performance ULSI," Proc. of the International Electron Devices Meeting, 241–242 (1995).
2. D. A. B. Miller, "Rationale and Challenges for Optical Interconnects to Electronic Chips," Proc. IEEE 88(6), 728–749 (2000).
3. R. Zia, J. A. Schuller, A. Chandran and M. L. Brongersma, "Plasmonics: the next chip-scale technology," Mater. Today 9(7–8), 20–27 (2006).
4. R. F. Oulton, V. J. Sorger, D. A. Genov, D. F. P. Pile, and X. Zhang, "A hybrid plasmonic waveguide for subwavelength confinement and long-range propagation," Nature Photon. 2(8), 496–500 (2008).
5. H. S. Chu, E. P. Li, P. Bai and R. Hegde, "Optical performance of single-mode hybrid dielectric-loaded plasmonic waveguide-based components," Appl. Phys. Lett. 96(22), 221103 (2010).
6. A.V. Krasavin and A. V. Zayats, "Silicon-based plasmonic waveguides," Opt. Express 18(11), 11791–11799 (2010).

7. P. Bai, M. X. Gu, X. C. Wei and E. P. Li, "Electrical detection of plasmonic waves using an ultra-compact structure via a nanocavity," Opt. Express 17(26), 24349–24357 (2009).
8. K. J. A. Ooi, P. Bai, M. X. Gu and L. K. Ang, "Design of a monopole-antenna-based resonant nanocavity for detection of optical power from hybrid plasmonic waveguides," Opt. Express 19(18), 17075–17085 (2011).
9. K. J. A. Ooi, P. Bai, M. X. Gu and L. K. Ang, "Plasmonic coupled-cavity system for enhancement of surface plasmon localization in plasmonic detectors," Nanotechnology 23(27), 275201 (2012).
10. K. F. MacDonald and N. I. Zheludev, "Active plasmonics: current status," Laser Photon. Rev. 4(4), 562–567 (2010).
11. T. Nikolajsen, K. Leosson and S. I. Bozhevolnyi, "Surface plasmon polariton based modulators and switches operating at telecom wavelengths," Appl. Phys. Lett. 85(24), 5833 (2004).
12. W. Cai, J. S. White, and M. L. Brongersma, "Compact, High-Speed and Power-Efficient Electrooptic Plasmonic Modulators," Nano Lett. 9(12), 4403–4411 (2009).
13. A. V. Krasavin and A. V. Zayats, "Electro-optic switching element for dielectric-loaded surface plasmon polariton waveguides," Appl. Phys. Lett. 97(4), 041107 (2010).
14. A.V. Krasavin and A.V. Zayats, "Photonic Signal Processing on Electronic Scales: Electro-Optical Field-Effect Nanoplasmonic Modulator," Phys. Rev. Lett. 109(5), 053901 (2012).
15. R. A. Soref and B. R. Bennett, "Electrooptical Effects in Silicon," IEEE J. Quantum Electron. 23(1), 123–129(1987).
16. M. J. Dicken, L. A. Sweatlock, D. Pacifici, H. J. Lezec, K. Bhattacharya, and H. A. Atwater, "Electrooptic Modulation in Thin Film Barium Titanate Plasmonic Interferometers," Nano Lett. 8(11), 4048–4052 (2008).
17. E. Feigenbaum, K. Diest and H. A. Atwater, "Unity-Order Index Change in Transparent Conducting Oxides at Visible Frequencies," Nano Lett. 10(6), 211 –2116 (2010).
18. V. J. Sorger, N. D. Lanzillotti-Kimura, R.-M. Ma and X. Zhang, "Ultra-compact silicon nanophotonic modulator with broadband response," Nanophotonics 1(1), 17–22 (2012).
19. N. F. Mott, "Metal-insulator transition," Rev. Mod. Phys. 40(4), 677–683 (1968).
20. H. W. Verleur, A. S. Barker, Jr., and C. N. Berglund, "Optical properties of $VO_2$ between 0.25 and 5 eV," Phys. Rev. 172(3), 788–798 (1968).
21. M. M. Qazilbash, M. Brehm, B. G. Chae, P. C. Ho, G. O. Andreev, et al., "Mott transition in $VO_2$ revealed by infrared spectroscopy and nano-imaging," Science 318(5857), 1750–1753 (2007).
22. M. F. Becker, A. B. Buckman, R. M. Walser, T. Lepine, P. Georges and A. Brun, "Femtosecond laser excitation of the semiconductor-metal phase-transition in $VO_2$," Appl. Phys. Lett. 65(12), 1507–1509 (1994).
23. B. Wu, A. Zimmers, H. Aubin, R. Ghosh, Y. Liu, and R. Lopez, "Electric-field-driven phase transition in vanadium dioxide," Phys. Rev. B 84(24), 241410(R) (2011).
24. J. Wei, H. Ji, W. Guo, A. H. Nevidomskyy and D. Natelson, "Hydrogen stabilization of metallic vanadium dioxide in single-crystal nanobeams," Nature Nanotech. 7(6), 357–362 (2012).
25. J. Cao, E. Ertekin, V. Srinivasan, W. Fan, S. Huang, et al., "Strain engineering and one-dimensional organization of metal-insulator domains in single-crystal vanadium dioxide beams," Nature Nanotech. 4(11), 732–737 (2009).

26. Z. Yang, C. Ko, and S. Ramanathan, "Oxide Electronics Utilizing Ultrafast Metal-Insulator Transitions," Annu. Rev. Mater. Res. 41(1), 337–367 (2011).
27. R. M. Briggs, I. M. Pryce and H. A. Atwater, "Compact silicon photonic waveguide modulator based on the vanadium dioxide metal-insulator phase transition," Opt. Express 18(11), 11192–11201 (2010).
28. L. A. Sweatlock and K. Diest, "Vanadium dioxide based plasmonic modulators," Opt. Express 20(8), 8700–8709 (2012).
29. S. Roberts, "Optical properties of copper," Phys. Rev. 118(6), 1509–1518 (1960).
30. C. S. T. Microwave Studio, 2012, http://www.cst.com/.
31. Z. Yang, C. Ko, V. Balakrishnan, G. Gopalakrishnan, and S. Ramanathan, "Dielectric and carrier transport properties of vanadium dioxide thin films across the phase transition utilizing gated capacitor devices," Phys. Rev. B 82, 205101 (2010).
32. E. E. Chain, "Optical properties of vanadium dioxide and vanadium pentoxide thin films," Appl. Opt. 30(19), 2782–2787 (1991).
33. S. Zhu, T. Y. Liow, G. Q. Lo, and D. L. Kwong, "Silicon-based horizontal nanoplasmonic slot waveguides for on-chip integration," Opt. Express 19(9), 888 –8902 (2011).
34. S. Zhu, T. Y. Liow, G. Q. Lo, and D. L. Kwong, "Fully complementary metal-oxide-semiconductor compatible nanoplasmonic slot waveguides for silicon electronic photonic integrated circuits," Appl. Phys. Lett. 98(2), 021107 (2011).
35. B. A. Kruger, A. Joushaghani, and J. K. S. Poon, "Design of electrically driven hybrid vanadium dioxide ($VO_2$) plasmonic switches," Opt. Express 20(21), 23598–23609 (2012).

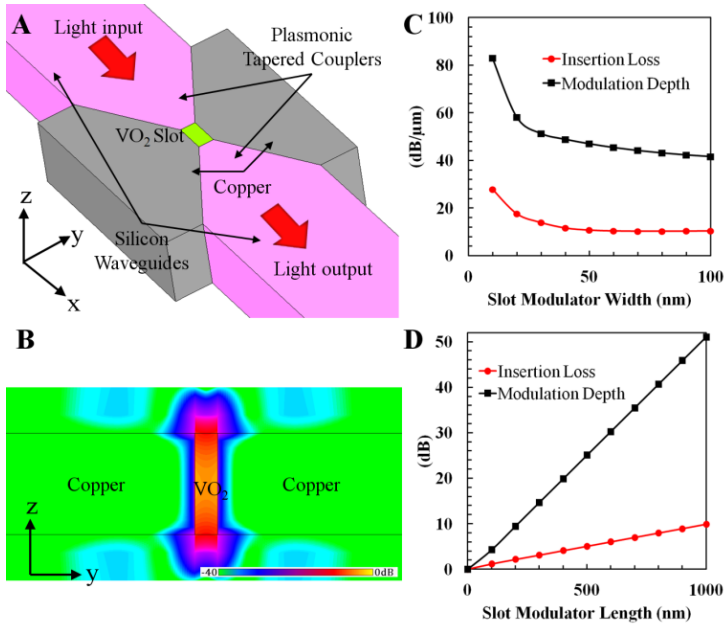

Figure 1 (A) Vanadium dioxide plasmonic slot waveguide modulator. (B) Electric-field map of surface plasmons confined in the Cu–$VO_2$–Cu slot, as seen from the cross-sectional cutting plane. (C) Variation of insertion loss and modulation depth with slot width. (D) Insertion loss and modulation depth for a 50nm wide slot waveguide modulator.

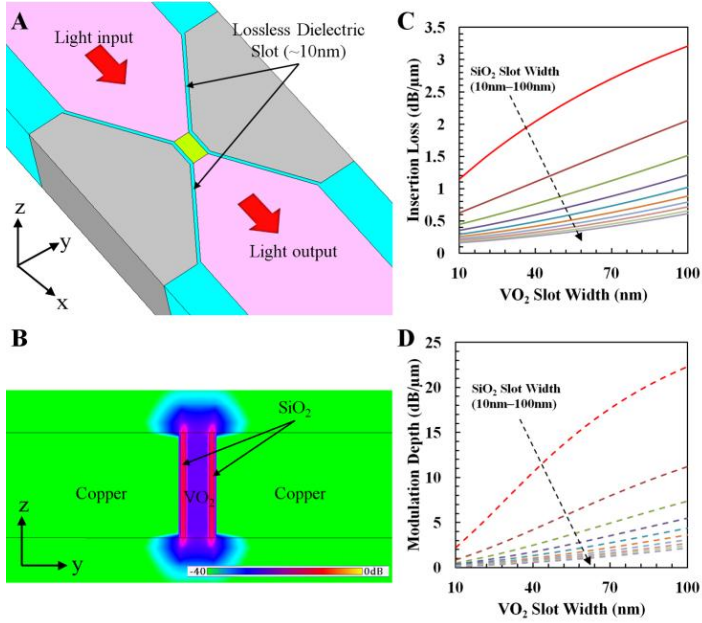

Figure 2 (A) Vanadium dioxide dual-mode plasmonic waveguide modulator. (B) Electric-field map showing strong confinement of surface plasmons in the Cu–$SiO_2$–$VO_2$ slots in the "on" state, as observed at cross-sectional cutting plane. (C) Variation of insertion loss with $VO_2$ and $SiO_2$ slot widths. (D) Variation of modulation depth with $VO_2$ and $SiO_2$ slot widths.

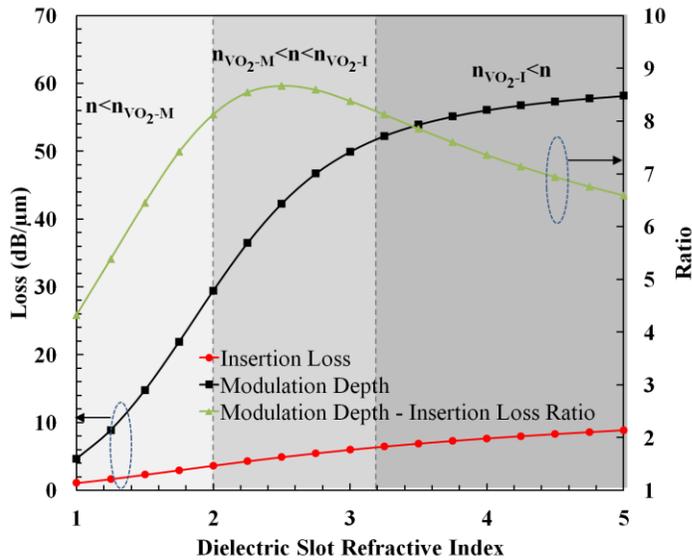

Figure 3 Left axis: insertion loss and modulation depth of the dual-mode plasmonic waveguide modulator as functions of refractive indices of the dielectric slot. Right axis: corresponding modulation depth to insertion loss ratio.

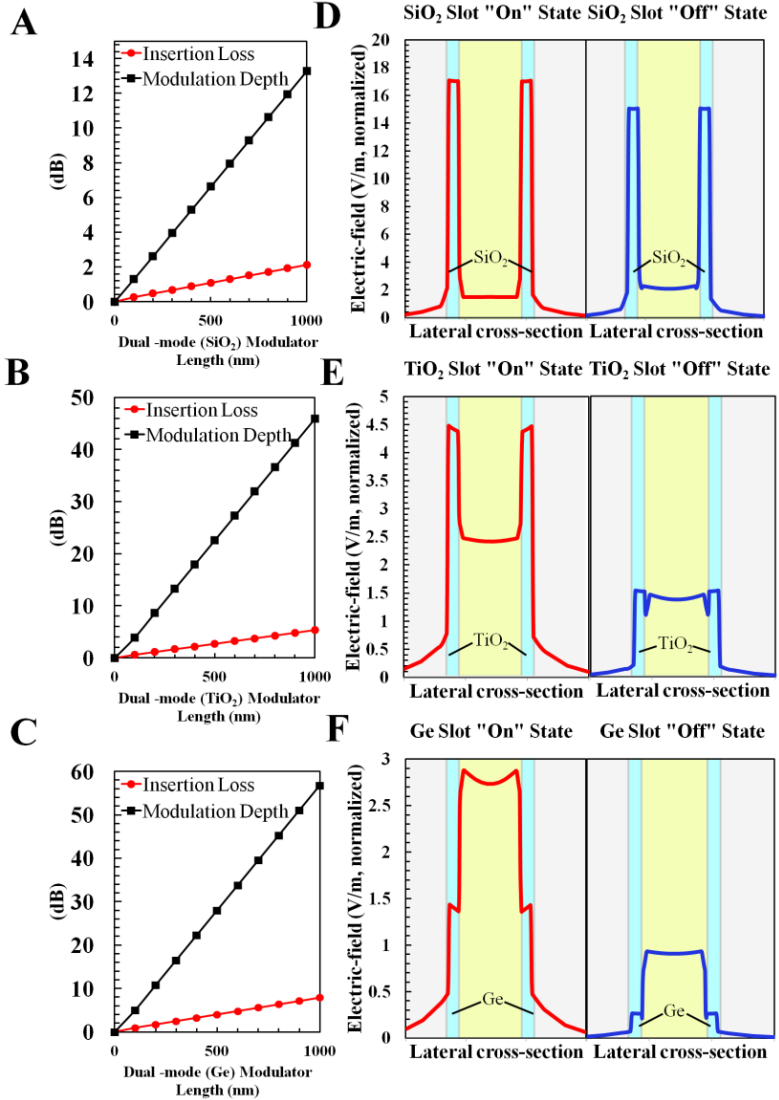

Figure 4 Insertion loss and modulation depth for the dual-mode plasmonic waveguide modulator with the dielectric slot filled with (A) $SiO_2$, (B) $TiO_2$ and (C) Ge material. (D), (E) and (F) are corresponding lateral electric-field profiles (normalized to refractive index) for the "on" and "off" states for $SiO_2$, $TiO_2$ and Ge-filled slots respectively.